\RequirePackage{lineno}
\documentclass[1p,times,procedia]{elsarticle}
\usepackage{nupha_ecrc}

\volume{00}

\firstpage{1}

\journalname{Nuclear Physics A}

\runauth{}

\jid{nupha}

\jnltitlelogo{Nuclear Physics A}

\usepackage{amssymb}

\usepackage[figuresright]{rotating}
\usepackage[export]{adjustbox}
\usepackage{wrapfig}

\newcommand{\pT}{$p_{\rm{\it{T}}}$}

\newcommand{\DirPho}{$\gamma_{dir}$}
\newcommand{\piZro}{$\pi^{0}$}
\newcommand{\GammaRich}{$\gamma_{rich}$}
\newcommand{\piZroRich}{$\pi^{0}_{rich}$}
\newcommand{\ptAssoc}{$p_{\rm{\it{T}}}^{assoc}$}
\newcommand{\ptTrig}{$p_{\rm{\it{T}}}^{trig}$}
\newcommand{\zT}{$z_{\rm{\it{T}}}$}

\newcommand{\IAApiZro}{$I_{AA}^{\pi^{0}}$}
\newcommand{\IAAg}{$I_{AA}^{\gamma_{dir}}$}
\begin{document}

\begin{frontmatter}



\dochead{}

\title{Direct-photon+hadron correlations to study parton
energy loss with the STAR
 experiment}


\author{Nihar Ranjan Sahoo (for the STAR collaboration)}

\address{Cyclotron Institute, Texas A$\&$M University, USA\\ Email id: nihar@rcf.rhic.bnl.gov}
\begin{abstract}
 We report new results of $\gamma_{dir}$+hadron and $\pi^{0}$+hadron azimuthal
  correlations as a measure of the away-side jet-like  correlated yields in 
  central Au+Au and p+p collisions at $\sqrt{s_{NN}}$ = 200 GeV in
  the STAR experiment from years 2011 and 2009 of data taking, respectively. The charged-hadron per-trigger yields at
mid-rapidity $(|\eta| < 1)$ and for transverse momenta 
$p_{T}^{assoc} > 1.2$ ~GeV/$c$
associated with $\gamma_{dir}$~ and $\pi^{0}$~ 
(for triggers $|\eta| < $0.9, 12 $< p_{T}^{trig} <$ 20 GeV/$c$) in central Au+Au
collisions are compared with p+p collisions. 
The $z_{T}$ ($= \frac{p_{T}^{assoc}}{p_{T}^{trig}}$) dependence, now extending down to $z_{T}$=0.1, of the suppression of the away-side associated yields is presented.
We observed that the suppression of away-side associated yields tends to vanish at lower \zT.
The dependence of the suppression on both $p_{T}^{assoc}$ and $p_{T}^{trig}$ is also discussed.
Finally, these results are compared with various model predictions.

\end{abstract}


\end{frontmatter}


\section{Introduction}
\label{intro}
The $\gamma_{dir}$+jet analysis has been discussed in many theoretical studies~\cite{Wang_Zhu,Wang,Renk,Wang_Huang_Sarcevic} to study parton energy loss as a probe of QCD matter at extreme temperature and pressure. 
The azimuthal correlations of the charged hadron with a direct
photon (\DirPho) trigger is considered as a promising probe in heavy-ion collisions for the
study of parton energy loss and necessary to understand the jet-quenching
mechanisms~\cite{Wang_Huang_Sarcevic}. In comparison with $\gamma_{dir}$ triggers, $\pi^{0}$-triggered charged hadron correlations can reveal path length dependence of energy loss due to the difference in geometrical biases as well as the the color factor dependence.
From our previous measurement~\cite{STAR_GJet}, it is observed that suppression of hadrons is independent of whether it is associated with a $\gamma_{dir}$- or $\pi^{0}$-trigger. In this work, we have performed $\gamma_{dir}$-hadron and  $\pi^{0}$-hadron correlation study, extending our kinematic region to lower \zT$= \frac{p_{T}^{assoc}}{p_{T}^{trig}}$~(down to \zT~=0.1) in order to understand the behavior of the medium modification factor at low \zT.

\section{STAR Detector and Transverse Shower Profile (TSP)}
\label{star_tsp}
The Time Projection Chamber
(TPC) is the main charged-particle tracking detector~\cite{STAR_TPC} in the STAR detector system. The Barrel Electromagnetic Calorimeter (BEMC) ~\cite{STAR_BEMC} is used both for
triggering on the events and for providing an energy measurement for the trigger photon or \piZro~in $|\eta|\leq 1.0$. 
STAR provides full 2$\pi$ azimuthal coverage and wide pseudo-rapidity ($|\eta|<1.0$) coverage. The Barrel Shower Maximum Detector (BSMD) provides high spatial resolution (both in the $\eta$ and $\phi$ planes). Detailed discussion about the BSMD can be found in Ref~\cite{STAR_GJet}. The data were taken by the STAR experiment and amount to an integrated luminosity of 2.8~nb$^{-1}$ of Au+Au collisions and 23~$ pb^{-1}$ of p+p collisions.

\begin{wrapfigure}{r}{0.5\textwidth}

\vspace{-15pt}
\includegraphics[width=0.8\linewidth]{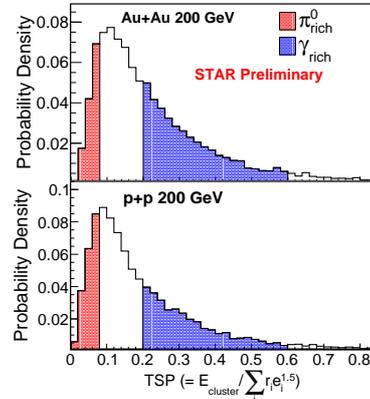} 
\vspace{-20pt}
\caption{The probability distribution for TSP for Au+Au (upper panel) at central $0-12\%$ collisions and p+p collisions (lower panel) are shown. The red and blue shaded regions represent the $\pi^{0}_{rich}$ and $\gamma_{rich}$ samples, respectively.}
 \vspace{-5pt}
\label{Fig1}
\end{wrapfigure}

In order to distinguish \piZro, which at high \pT~ decay to two photons that are close together, from single $\gamma$-clusters, a transverse shower-shape analysis is performed. In this method, the overall BEMC cluster energy ($E_{cluster}$), the individual BSMD strip energies ($e_{i}$), and the distances of the strips ($r_{i}$) from the center of the cluster are used. To quantify the shower shape, an observable we call the ``Transverse Shower Profile'' (TSP $=\frac{E_{cluster}}{\sum_{i}e_{i}r_{i}^{1.5}}$) is used.  
The \piZroRich~(nearly pure sample of \piZro) and \GammaRich~(enhanced fraction of
\DirPho) samples are selected by requiring TSP $< 0.08$ and $0.2< $TSP$<0.6$, respectively, in both p+p and Au+Au collisions as shown in Fig.~\ref{Fig1}. These TSP conditions are decided based on simulation study.  
The associated charged particles are selected in range $1.2~$GeV/$c<$ \ptAssoc~within $|\eta|<1.0$, whereas $\pi^{0}$ and $\gamma_{dir}$ are triggered within $12 <$ \ptTrig~$< 20~$GeV/$c$. The away-side associated charged hadron yield for direct-photon trigger, assuming zero near-side $\gamma_{dir}$ yield, is estimated using the following expression   
$ Y_{\gamma_{dir}-h} = \frac{Y^{a}_{\gamma_{rich}-h}-RY^{a}_{\pi^{0}-h}}{1-R}$.
Here $Y^{a}_{\gamma_{rich}-h}$ ($Y^{a}_{\pi^{0}-h}$) 
represents the away-side yield of \GammaRich~(\piZro); and
$R = \frac{Y^{n}_{\gamma_{rich}-h}}{Y^{n}_{\pi^{0}-h}}$, the ratio
of near-side yield in the $\gamma_{rich}$-triggered correlation function
to the  near-side yield in the $\pi^{0}$-triggered correlation function.
Then $1-R= \frac{N^{\gamma^{dir}}}{N^{\gamma_{rich}}}$, where
$N^{\gamma^{dir}}$ ($N^{\gamma_{rich}}$) is the
number of \DirPho~(\GammaRich) triggers. The values of $1-R$, representing 
the fraction of $\gamma^{dir}$ in the $\gamma^{rich}$ trigger sample, are found
to be 40$\%$ and 70$\%$ for p+p and the central Au+Au
collisions. From the azimuthal angular correlation functions of $\gamma_{rich}$- and $\pi^{0}_{rich}$-triggered associated charged hadrons, away-side and near-side charged hadron yields are calculated after background subtraction and the pair acceptance correction. In Au+Au collisions, the background level is modulated with an elliptic flow component in the azimuthal correlations.
\section{Results and Discussion}
\begin{figure*}
\centering
\begin{minipage}{.5\textwidth}
  \centering
  \includegraphics[width=0.8\linewidth]{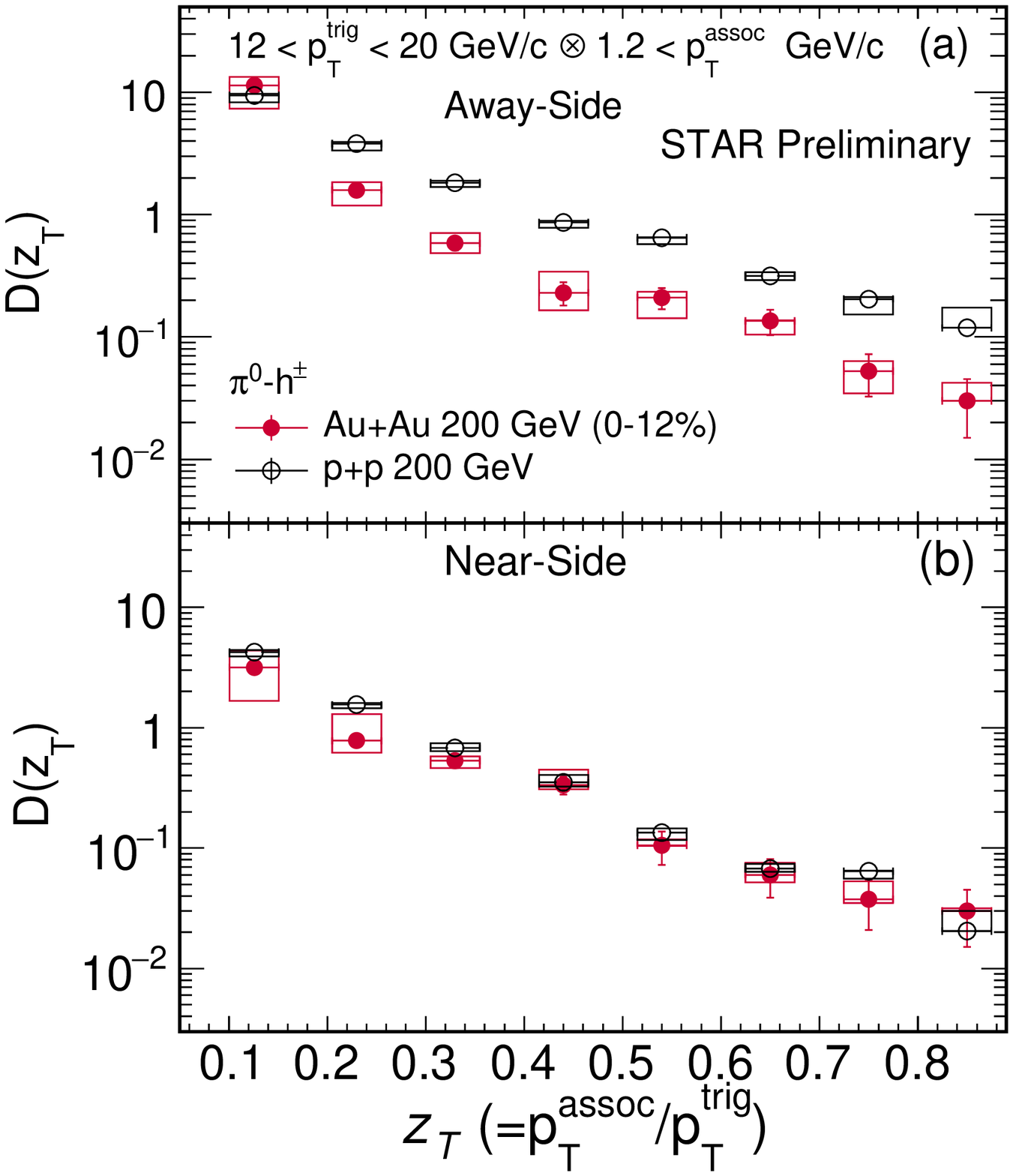}
\end{minipage}%
\begin{minipage}{.5\textwidth}
  \centering
  \includegraphics[width=0.8\linewidth]{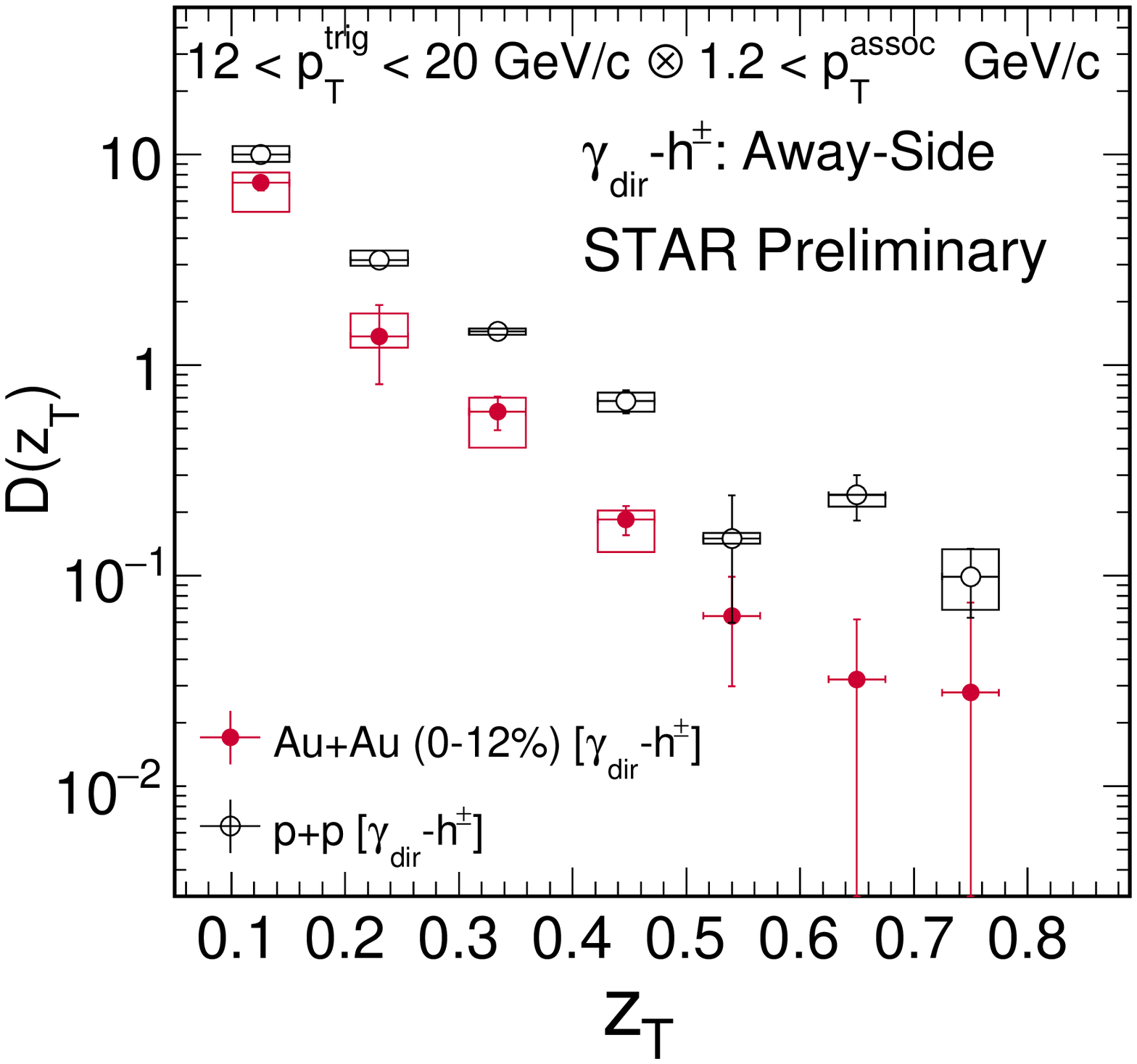}
\end{minipage}
\vspace{-5pt}
 \caption{Left panel: The $z_{T}$ dependence of \piZro-$h^{\pm}$ away-side (upper) and near-side (lower) associated charged-hadron yields per trigger for Au+Au at 0-12$\%$ centrality  (filled symbols) and  p+p (open symbols) collisions in left panel. Right panel: The \DirPho-$h^{\pm}$ away-side associated charged-hadron yields per  trigger for Au+Au at 0-12$\%$ centrality (filled circles) and  p+p (open circles) collisions. Vertical lines represent the statistical  errors, and the boxes represent systematic uncertainties.}
  \label{Fig2}
\end{figure*}

\begin{figure}
\begin{center}

\includegraphics[width=0.5\linewidth]{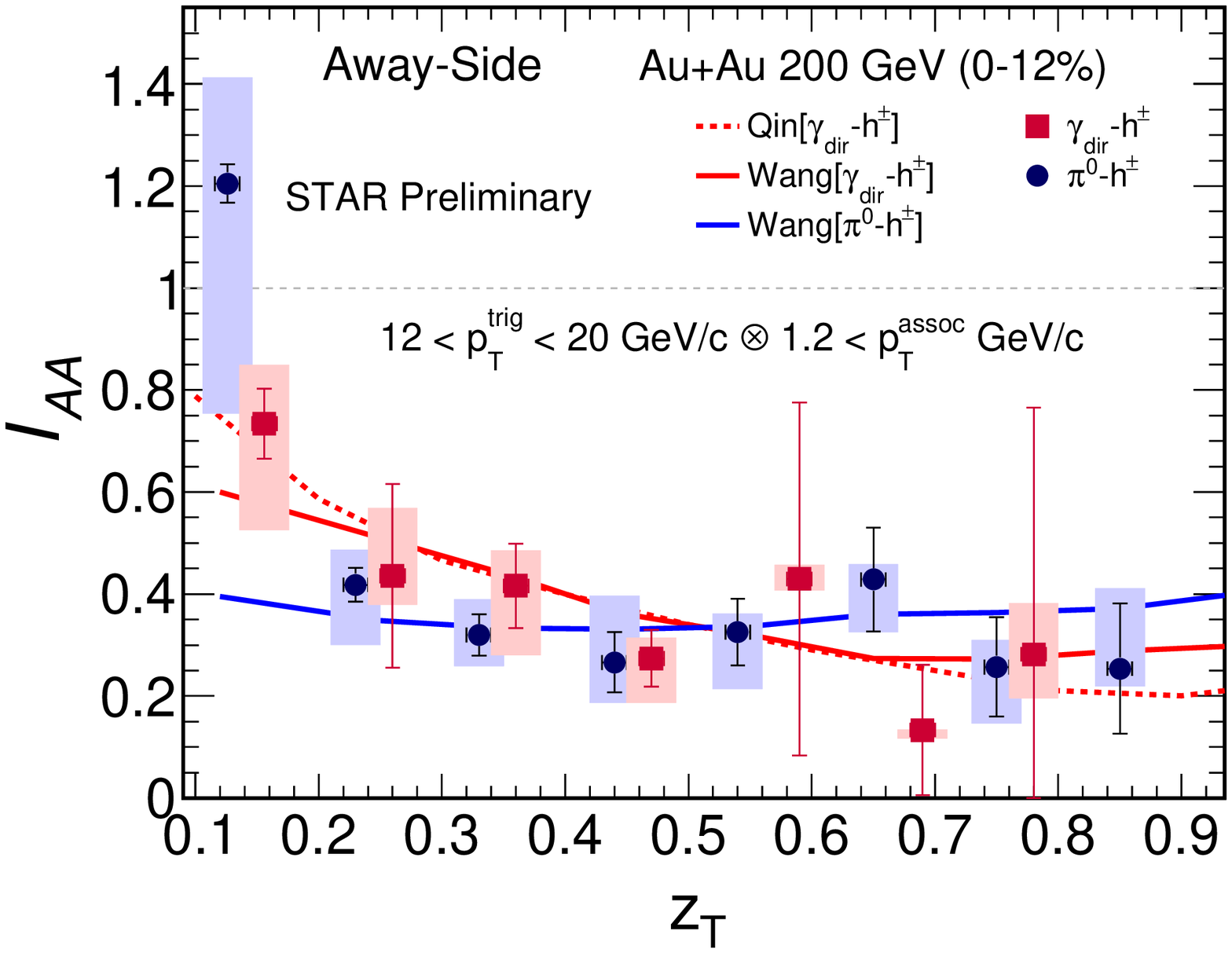} 
\caption{The  \IAAg~(red
  squares) and \IAApiZro~triggers (blue circles) are plotted as a function of \zT$=p_T^{assoc}/p_T^{trig}$. The points for \IAAg~ are shifted by $+0.03$ in $z_T$ for visibility.  The vertical lines represent statistical
  error bars and boxes represent systematic errors. The lines
  represents theoretical model predictions~\cite{Qin,Wang}. }
  \end{center}
\label{Fig3}
 \centering
  \includegraphics[width=0.5\linewidth]{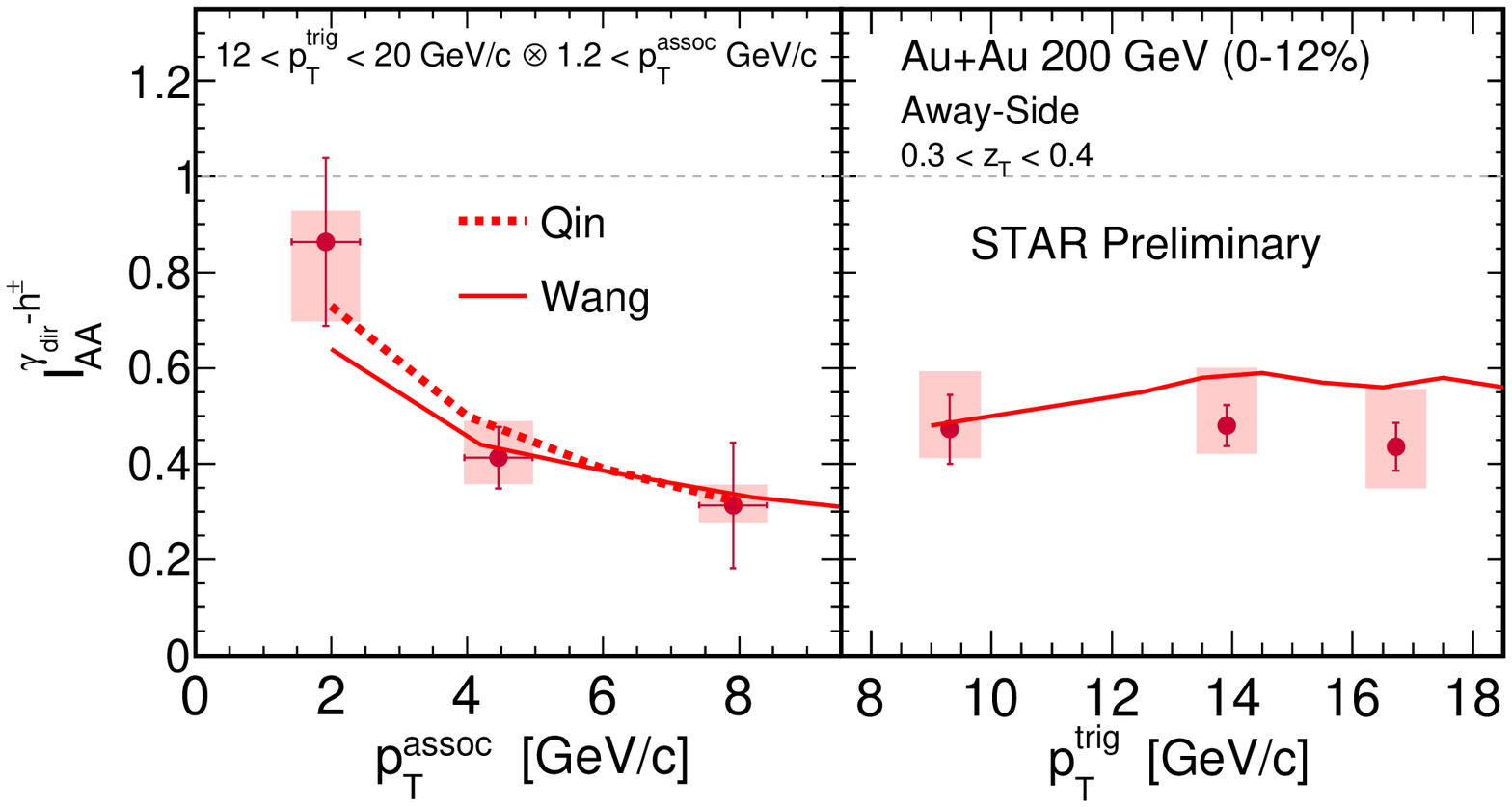}
\vspace{-10pt}
\caption{
  The values of \IAAg~ are plotted as a function of \ptAssoc~(left panel) and \ptTrig~(right panel). The solid line (broken line) represents Wang~\cite{Wang} (Qin~\cite{Qin})model prediction. The vertical line and shaded boxes represents statistical and systematic errors, respectively.}
\vspace{-10pt}
  \label{Fig4} 

\end{figure}

The away- ($|\Delta\phi-\pi|<1.4$) and near-side ($|\Delta\phi|<1.4$) charged hadron associated yields for $\pi^{0}$ triggered are plotted as a function of \zT~for Au+Au at 0-12$\%$ central and p+p collisions in Fig.~\ref{Fig2}. It is observed that away-side associated hadrons are highly suppressed at high \zT~whereas that of near-side show no suppression in Au+Au relative to p+p collisions. 
The away-side $D(z_{T})$ for $\gamma_{dir}$ triggers as a function of \zT~ for central Au+Au and p+p collisions are plotted in Fig.~\ref{Fig2} (right panel), which shows the associated yields are suppressed more at high-\zT~compared with low-\zT. 

The medium modification for $\gamma_{dir}-$ and $\pi^{0}-$triggered recoil jet production as a function of \zT~is defined as  $I_{AA}~=~D(z_{T})^{AA}/D(z_{T})^{pp}$, of the per-trigger conditional yields in Au+Au to those in p+p collisions. The away-side medium modification factor for $\gamma_{dir}$ (\IAAg) and $\pi^{0}$ (\IAApiZro)-triggers as a function of \zT~are plotted in Fig.~\ref{Fig3}. At low \zT~(0.1~ $<$ \zT $<$ ~0.3), both \IAAg~and \IAApiZro~show less suppression than at high \zT. \IAApiZro~and \IAAg~show similar suppression within uncertainties. The theoretical model predictions, labeled as Qin~\cite{Qin} and Wang~\cite{Wang}, using the same kinematic coverage for $\gamma_{dir}$-tagged away-side charged hadron yields are compared with the data. Neither model includes a redistribution of the lost energy to the lower \pT~jet fragments. 
The \IAAg~are plotted as a function of \ptAssoc~and \ptTrig (for 0.3 $<$ \zT$< $0.4)~in Fig.~\ref{Fig4}. It suggests that at low-\ptAssoc~hadrons on the away-side are not as suppressed as those at high \ptAssoc, whereas the away-side parton energy loss is less sensitive to the initial parton energy at \ptTrig~range of 8 to 20 GeV/$c$. Both models
predict the data quite well.
    
\section{Summary}
\vspace{-5pt}
The correlation studies of $\gamma_{dir}$-hadron and $\pi^{0}$-hadron are performed to understand the effect of parton energy loss in the medium formed in Au+Au, at 0-12$\%$ central collisions by comparing Au+Au with p+p collisions at $\surd{s_{NN}}=$200 GeV. It is observed that \IAAg~ and \IAApiZro~show similar suppression. At low \zT, both \IAAg~ and \IAApiZro~results show less suppression compared with high~\zT. 
\IAAg~doesn't show dependence on the initial parton energy in the range of 8$<$\ptTrig$<$20 GeV/$c$. 
Two model predictions are compared with the data,  and neither model includes redistribution of the lost energy to the lower \pT~jet fragments.
 \vspace{-5pt}
\section*{Acknowledgments}
 \vspace{-10pt}
This conference proceedings is supported by the US Department of
Energy under the grant DE-FG02-07ER41485. We thank X. N. Wang and G.-Y Qin for providing their model predictions.





\bibliographystyle{elsarticle-num}
\bibliography{<your-bib-database>}



\end{document}